\documentclass[12pt]{article}
\usepackage{color}
\usepackage{graphicx}
\usepackage{rotating}
\def\nabstar#1{\nabla\kern-0.5pt\smash{\raise 4.5pt\hbox{$\ast$}}
               \kern-4.5pt_{#1}}

\def\drvstar#1{\partial\kern-0.5pt\smash{\raise 4.5pt\hbox{$\ast$}}
               \kern-5.0pt_{#1}}

\def\newline{\relax\ifhmode\null\hfil\break\else\nonhmodeerr@\newline\fi}
\def\frac#1#2{{#1\over#2}}
\def\text#1{{\hbox{\rm #1}}}
\def\flushpar{{\par \noindent}}

\newcommand{\beq}{\begin{equation}}
\newcommand{\eeq}{\end{equation}}
\newcommand{\bea}{\begin{eqnarray}}
\newcommand{\eea}{\end{eqnarray}}
\def\Id{ \mbox{1\hspace{-1.2mm}I} }
\def\EQ{\hspace{-2mm} &=& \hspace{-2mm}}

\def\BA{\begin{eqnarray}}
\def\EA{\end{eqnarray}}
\def\BAN{\begin{eqnarray*}}
\def\EAN{\end{eqnarray*}}

\def\tr{\mbox{tr}}

\def\gm5{\gamma_5}

\input epsf.sty
\newdimen\psfigsize
\def\psfigure#1 #2 #3 #4 #5{
    \begin{figure}[tbh]
      \begin{center}
      \vbox{
        \null\vskip-0.2in\hskip#2
        \epsfxsize=#1
        \epsfbox{#4}
        \vskip -0.3in
        \caption {#5 \label{#3}}
        \vskip 0.0 true in plus 0.3 true in
      }
      \end{center}
   \end{figure}
}
\usepackage{graphicx}
\begin{document}
\thispagestyle{empty}
\begin{flushright}
NTUTH-04-505A (v5)\\
June 2005 \\
\end{flushright}
\bigskip\bigskip\bigskip
\vskip 2.5truecm
\begin{center}
{\LARGE {Study of $ \Theta^+(udud \bar s) $ in lattice QCD \\
         with exact chiral symmetry
         }}
\end{center}
\vskip 1.0truecm
\centerline{Ting-Wai Chiu and Tung-Han Hsieh}
\vskip5mm
\centerline{Department of Physics and} 
\centerline{National Center for Theoretical Sciences at Taipei,} 
\centerline{National Taiwan University, Taipei 106, Taiwan}
\vskip 1cm
\bigskip \nopagebreak \begin{abstract}
\noindent

We investigate the mass spectrum of the pentaquark baryon ($ udud \bar s $) 
in quenched lattice QCD with exact chiral symmetry.  
Using 3 different interpolating operators for $ \Theta $,  
we measure the $ 3 \times 3 $ correlation matrix  
and obtain the eigenvalues $ A^{\pm} (t) $ with $ \pm $ parity,   
for 100 gauge configurations generated with Wilson gauge 
action at $ \beta = 6.1 $ on the $ 20^3 \times 40 $ lattice. 
For the lowest-lying $ J^P = 1/2^- $ state,
its mass is almost identical to that of the KN s-wave,
while for the lowest-lying $ J^P = 1/2^+ $ state,
its mass is smaller than that of the
KN p-wave, especially for the regime $ m_u < m_s $.
By chiral extrapolation (linear in $m_\pi^2$) to $ m_\pi = 135 $ MeV,
we obtain the masses of the lowest-lying states:
$ m(1/2^-) = 1424(57) $ MeV and $ m(1/2^+) = 1562(121) $ MeV,
in agreement with the masses of $ m_K + m_N \simeq 1430 $ MeV and
$ \Theta^+(1540) $ respectively.



\end{abstract}
\vskip 1.5cm 
\newpage\setcounter{page}1

\section{Introduction}

The recent experimental observation of the exotic baryon $ \Theta^+(1540) $
(with the quantum numbers of $ K^+ n $)
by LEPS collaboration\cite{Nakano:2003qx}
at Spring-8 and the subsequent confirmation
\cite{Barmin:2003vv,Stepanyan:2003qr,Barth:2003es,Asratyan:2003cb,
Kubarovsky:2003fi,Airapetian:2003ri,Aleev:2004sa}
from some experimental groups has become one of the most interesting topics
in hadron physics. The remarkable features of $ \Theta^+(1540) $ are its 
strangeness $ S=+1 $, and its exceptionally narrow decay width 
($ < 15 $ MeV) even though it is $ \sim 100 $ MeV above the KN threshold.
Its strangeness $ S = +1 $ immediately implies that it cannot be an 
ordinary baryon composed of three quarks. Its minimal quark content 
is $ u d u d \bar s $.
Nevertheless, there are quite a number of experiments\cite{Hicks:2004ge}
which so far have {\it not} observed $ \Theta^+(1540)$ or any pentaquarks.
This casts some doubts about the existence of $ \Theta^+(1540) $.

Historically, the experimental search for $ \Theta^+(1540) $ was motivated
by the predictions of the chiral-soliton model\cite{Diakonov:1997mm},
an outgrowth of the Skyrme model\cite{csm}.
Even though the chiral solition model
seems to provide very close predictions for the mass and the width of
$ \Theta^+(1540) $, obviously, it cannot reproduce all aspects of QCD,
the fundamental theory of strong interactions.
Now {\it the central theoretical question is whether the spectrum of QCD
possesses $ \Theta^+ $ with the correct quantum numbers, mass,
and decay width.}

At present, the most viable approach to solve QCD nonperturbatively
from the first principles is lattice QCD. 
Explicitly, one needs to construct an 
interpolating operator which has a significant overlap with 
the pentaquark baryon states. Then one computes the time-correlation
function of this interpolating operator, and from which to extract
the masses of its even and odd parity states respectively. 
However, any $ (udud\bar s) $ operator must couple to hadronic states 
with the same quantum numbers (e.g., $ KN $ scattering states). 
It is necessary to disentangle the lowest-lying pentaquark states 
from the $ KN $ scattering states, as well as the excited pentaquark states.

To this end, we adopt the so-called variational method
\cite{Michael:1985ne,Luscher:1990ck},
and use three different interpolating operators
for $ \Theta(udud \bar s) $, to compute their $ 3 \times 3 $ correlation
matrix, and from its eigenvalues to extract the masses of the
even and odd parity states. This is the first lattice QCD study of 
$ \Theta^+ $ with $ 3 \times 3 $ correlation matrix.
These three interpolating operators (with $ I=0 $) are:   
\bea
\label{eq:O1}
{(O_1)}_{x\alpha} \EQ
  [{\bf u}^T C \gamma_5 {\bf d}]_{xc} \
\{ \bar{\bf s}_{x \beta e} (\gamma_5)_{\beta\eta} {\bf u}_{x \eta e}
 (\gamma_5 {\bf d})_{x \alpha c}
- \bar{\bf s}_{x \beta e} (\gamma_5)_{\beta\eta} {\bf d}_{x \eta e} 
(\gamma_5 {\bf u})_{x \alpha c} \}
\\
\label{eq:O2}
{(O_2)}_{x\alpha} \EQ
  [{\bf u}^T C \gamma_5 {\bf d}]_{xc} \
\{ \bar{\bf s}_{x \beta e} (\gamma_5)_{\beta\eta} {\bf u}_{x \eta c}
   (\gamma_5 {\bf d})_{x \alpha e}
- \bar{\bf s}_{x \beta e} (\gamma_5)_{\beta\eta} {\bf d}_{x \eta c} 
 (\gamma_5 {\bf u})_{x \alpha e} \}   \\
\label{eq:O3}
{(O_3)}_{x\alpha} \EQ \epsilon_{cde}
  [{\bf u}^T C \gamma_5 {\bf d}]_{xc} \
  [{\bf u}^T C {\bf d}]_{xd} \ (C \bar {\bf s}^T)_{x\alpha e}
\eea
where ${\bf u}$, ${\bf d}$ and ${\bf s}$ denote the quark fields;
$ \epsilon_{cde} $ is the completely antisymmetric tensor;
$ x $, $ \{ c,d,e \} $ and $ \{ \alpha, \beta, \eta \} $ 
denote the lattice site, color, and Dirac indices respectively;
and $ C $ is the charge conjugation operator satisfying 
{$ C \gamma_\mu C^{-1} = -\gamma_\mu^T $} and 
{$ (C \gamma_5)^T=-C\gamma_5 $}.    
Here the diquark operator is defined as 
\bea
[{\bf u}^T \Gamma {\bf d} ]_{xa} \equiv \epsilon_{abc} ( 
 {\bf u}_{x\alpha b} \Gamma_{\alpha\beta} {\bf d}_{x\beta c}
-{\bf d}_{x\alpha b} \Gamma_{\alpha\beta} {\bf u}_{x\beta c} )
\eea 
where $ \Gamma_{\alpha\beta} = -\Gamma_{\beta\alpha} $.
Thus the diquark transforms like a spin singlet ($1_s$), 
color anti-triplet ($ \bar 3_c $), 
and flavor anti-triplet ($ \bar 3_f $). For $ \Gamma = C \gamma_5 $, it
transforms as a scalar, while for $ \Gamma = C $, it transforms like 
a pseudoscalar. Here $ O_1 $, $ O_2 $, and $ O_3 $ all transform 
like an even operator under parity.

The operator $ O_1 $ is similar to the naive kaon $\otimes$ nucleon operator 
which was used by Mathur et al. \cite{Mathur:2004jr}.
The operator $ O_2 $ was first considered by Zhu \cite{Zhu:2003ba}
and was adopted by Csikor et al. \cite{Csikor:2003ng}
in their lattice study. The difference between $ O_1 $ and $ O_2 $ 
is that, in the latter case, the color index of the $ {\bf u} ({\bf d}) $ 
quark in the kaon is swapped with that of the ${\bf d} ({\bf u}) $ quark 
in the nucleon such that the ``kaon" and the ``nucleon" do not appear 
as color singlets. The operator $ O_3 $ is motivatied by the 
Jaffe-Wilzcek (diquark-diquark-antiquark) model \cite{Jaffe:2003sg}, 
which was considered by Sugiyama et al. \cite{Sugiyama:2003zk}, 
Sasaki \cite{Sasaki:2003gi}, and was adopted by  
Chiu and Hsieh \cite{Chiu:2004gg}, and Ishii et al. \cite{Ishii:2004qe}.

In the Jaffe-Wilzcek model, each pair of $ [u d] $ form a diquark 
which transforms like a spin singlet ($1_s$), 
color anti-triplet ($ \bar 3_c $), and flavor anti-triplet ($ \bar 3_f $). 
Then the pentaquark baryon $ \Theta([ud][ud] \bar s) $ emerges as the 
color singlet in 
%
 $  ( \bar 3_c \times \bar 3_c) \times \bar 3_c 
   = 1_c + 8_c + 8_c + \overline{10_c} $, 
and a member (with $S=+1$ and $I=0$) of the flavor anti-decuplet in 
%
 $ \bar 3_f \times \bar 3_f \times \bar 3_f   
  = 1_f + 8_f + 8_f + \overline{10}_f $. 
Now, if one attempts to construct a local interpolating operator
for $[ud][ud]\bar s $, then these two identical diquarks must be
chosen to transform differently (i.e., one scalar and one pseudoscalar),  
otherwise $ \epsilon_{abc} [ud]_{xb} [ud]_{xc} \bar s_{x\alpha a} $
is identically zero since diquarks are bosons.
Thus, when the orbital angular momentum of this
scalar-pseudoscalar-antifermion system is zero
(i.e., the lowest lying state), its parity is even rather than odd.
Alternatively, if these two diquarks are located at two different sites,
then both diquark operators can be chosen to be scalar,
however, they must be antisymmetric in space,  
i.e., with odd integer orbital angular momentum. 
Thus the parity of lowest lying state of
this scalar-scalar-antifermion system is even, as suggested in the
original Jaffe-Wilzcek model. (Note that all correlated quark models
e.g., Karliner-Lipkin model \cite{Karliner:2003dt} and
      flavor-spin model \cite{Stancu:2003if}, 
advocate that the parity of $ \Theta^+(1540) $ is positive.)

Evidently, the diquark operator plays an important role 
in constructing sources for pentaquark baryons as well as
3-quark baryons.   
The possibility of forming multiquark hadrons through diquark 
correlations was proposed by Jaffe in 1977 \cite{Jaffe:1977ig}.
Although the idea is essentially based on the color-spin 
interaction between the quarks (through one gluon exchange), 
its salient features seem to persist even at the hadronic distance 
scale where QCD is strongly coupled. Thus, it is interesting to see 
whether such multiquark hadrons (e.g., pentaquark baryons) do exist in the 
spectrum of QCD. 

In this paper, we use the optimal domain-wall fermion 
\cite{Chiu:2002ir} to study the pentaquark baryons. 
The salient features of optimal lattice domain-wall fermion are:
(i) The quark propagator as well as the effective 4D lattice 
Dirac operator for internal fermion loops have optimal chiral
symmetry for any $ N_s $ (number of sites in the 5-th dimension) 
and gauge background; 
(ii) The quark fields and hadron observables
manifest the discrete symmetries of their counterparts in continuum; 
(iii) The quark action is ultralocal on the 5 dimensional lattice, 
thus the dynamical quark can be simulated with the standard 
hybrid monte carlo algorithm;  
(iv) The quark propagator in gauge background can be computed 
efficiently through the effective 4D lattice Dirac operator.

\begin{sidewaystable}
\begin{center}
\begin{tabular}{c|ccccccccccc}
Ref.  & Operator 
      & Quark & Gauge 
      & a  &  L  
      & $ m_{\pi}^{min} $ 
      & $ N(m_u < m_s)$ 
      & $ m(1/2^-)$, S/R & $ m(1/2^+) $, S/R 
      & Signal/Parity
\\
\hline
\cite{Csikor:2003ng}
   &  $ O_1 + \alpha O_2 $   
   &  W & W 
   &  0.09 & 1.8  
   &  420 
   &  3   
   &  1539(50), R & 2710(79) 
   &  Yes/$-$  
\\
\cite{Sasaki:2003gi}
   &  $ O_3 $                 
   &  W & W 
   &  0.07 & 2.2 
   &  650 
   &  1   
   &  1840(80), R & 2940(130)  
   &  Yes/$-$ 
\\
This work 
   &  $ \{O_{1,2,3} \}_{3\times 3} $ 
   &  Odwf & W 
   &  0.09 & 1.8
   &  440 
   &  10 
   &  1433(72), S & 1562(121), R   
   &  Yes/$+$ 
\\
\cite{Mathur:2004jr}
   &  $ O_1 $ 
   &  Ov & Iw 
   &  0.20  & 2.4/3.2 
   &  180  
   &  13 
   &  $\sim 1450 $, S & $ \sim 1650 $, S 
   &  No  
\\
\cite{Ishii:2004qe}
   &  $ O_3 $ 
   &  W & W    
   &  0.18  & 2.1 
   &  656  
   &  0 
   &  1750(40), S & 2250(120) 
   &  No  
\\
\hline
\end{tabular}
\end{center}
\caption{Summary of current lattice QCD results for $ \Theta^+ $.
Here all masses are in units of MeV, the lattice spacing (a) and the box size
(L) are in units of Fermi. The notations for the symbols are: 
W -- Wilson fermion or Wilson gauge action; Ov -- overlap fermion; 
Iw -- Iwasaki gauge action; $ m_\pi^{min} $ -- the smallest pion mass; 
$ N(m_u < m_s) $ -- the number of data points satisfying the condition 
$ m_u < m_s $; $ m(1/2^\pm) $ -- the mass of $ 1/2^\pm $ 
state via chiral extrapolation; S/R -- Scattering state/Resonance. 
There are two lattice studies \cite{Alexandrou:2004ws,Takahashi:2004sc} 
which have not listed here, due to lack of the information for 
some of the entries.  
}
\label{tab:Theta_LQCD_0205}
\end{sidewaystable}

Before we turn to our results, it is instructive to review the current 
status of quenched lattice QCD for $ \Theta^+ $, 
as summarized in Table \ref{tab:Theta_LQCD_0205}. 
At first sight, current lattice QCD results for $ \Theta^+ $ 
seems to disagree with each other. Note that the different claims listed 
in the last column of Table \ref{tab:Theta_LQCD_0205} 
already cover all possible outcomes. 
Obviously, some of the claims in Table \ref{tab:Theta_LQCD_0205} 
cannot be sustained for a long time, no matter what is the
experimental outcome.         
Now if one compares the essential features among these 
{\it exploratory} lattice studies, one might understand what 
could be the causes for these different claims. 
In the following, we pinpoint the crucial features 
which may have direct impacts to these claims. 
 
So far, all lattice QCD simulations are performed at unphysically 
large $ m_u $, thus it is necessary to chirally extrapolate
to physical $ m_u $ (or equivalently $ m_\pi = 135 $ MeV).  
Then a crucial question is how good are the data points
used for chiral extrapolation, i.e., 
how many data points are obtained with $ m_u < m_s $ 
(an obviously physical condition ought to be satisfied), 
and what is the smallest $ m_u $ (or equivalently $ m_\pi $) 
in these data points. 
These two questions are answered in the columns with headings 
$ N(m_u < m_s) $ and $ m_\pi^{min} $ respectively. 
Obviously, if $ m_\pi^{min} $ is too large, and/or 
the number $ N(m_u < m_s) $ is too small, then the chiral 
extrapolation would tend to overestimate the masses, 
especially for the excited states.   

Another important question is whether the interpolating operator 
one uses has a significant overlap with the pentaquark state. 
If it has little overlap with the pentaquark state, 
then the signal might be too weak to be detected. 
As we will see below, $ O_1 $, $ O_2 $ and $ O_3 $
all have good overlap with the lowest-lying negative parity state
for the entire range of $ m_u $. 
However, the lowest-lying negative parity state turns out to be 
nothing but the $KN$ s-wave scattering state. 
On the other hand, for the positive parity channel,  
only $ O_3 $ has a significant overlap with the lowest-lying 
positive parity state, in the regime $ m_u \le m_s/2 $. 
As we will see below, this positive parity state is ruled 
out to be $ KN $ p-wave, $ KN^* $ s-wave
(where $ N^* $ is the lowest negative parity state of nucleon), 
or $ KN\eta' $ s-wave 
(where $\eta'$ is the artifact due to the quenched approximation). 

At the end of this paper, we will return to Table \ref{tab:Theta_LQCD_0205} 
to discuss what could be the causes for the different claims 
in these exploratory lattice studies. 

The outline of this paper is as follows. 
In Section 2, we outline our computatation of quark propagators. 
In Section 3, we outline our determination of the lattice spacing $ a $, 
and the strange quark bare mass $ m_s $.
In Section 4, we present our results of the masses of the 
even and odd parity states extracted from 
the $ 3 \times 3 $ correlation matrix of $ O_1 $, $ O_2 $ and $ O_3 $. 
These are the first lattice QCD results   
using $ 3 \times 3 $ correlators for $ \Theta^+ $.
In Section 5, we investigate the KN scattering states with
``disconnected" KN (i.e., without quark exchanges between K and N) 
correlation function, and use them to identify the KN scattering 
states in the spectrum of $ 3 \times 3 $ correlation matrix.
In Section 6, we discuss the current lattice QCD results for 
$ \Theta^+ $, and conclude with some remarks. 
In the Appendix, we include our results of the masses of 
the even and odd parity states extracted from the 
time-correlation functions of $ O_1 $, $ O_2 $, and $ O_3 $ respectively.

\section{Computation of quark propagators}

Now it is straightforward to work out the baryon propagator
$ \langle \Theta_{x\alpha} \bar\Theta_{y\delta} \rangle $ 
in terms of quark propagators. 
In lattice QCD with exact chiral symmetry,
quark propagator with bare mass $ m_q $ is of the form 
$ (D_c + m_q )^{-1} $ \cite{Chiu:1998eu}, 
where $ D_c $ is exactly chirally symmetric
at finite lattice spacing. 
In the continuum limit, $ (D_c + m_q)^{-1} $ reproduces
$ [ \gamma_\mu ( \partial_\mu + i A_\mu ) + m_q ]^{-1} $. 
For optimal domain-wall fermion with 
$ N_s + 2 $ sites in the fifth dimension,   
\BAN
D_c \EQ 2m_0 \frac{1 + \gamma_5 S(H_w)}{1 - \gamma_5 S(H_w)}, \\
S(H_w) \EQ \frac{1 - \prod_{s=1}^{N_s} T_s}
                {1 + \prod_{s=1}^{N_s} T_s}, \\
 T_s \EQ \frac{1 - \omega_s H_w }{1 + \omega_s H_w},  \hspace{4mm}
 H_w = \gamma_5 D_w,  
\EAN
where $ D_w $ is the standard Wilson Dirac operator plus a negative 
parameter $ -m_0 $ ($0<m_0<2$), and $ \{ \omega_s \} $ are a set of 
weights specified by an exact formula such that $ D_c $ 
possesses the optimal chiral symmetry \cite{Chiu:2002ir}. Since 
\BAN
( D_c + m_q )^{-1} 
= (1-rm_q)^{-1} [ D^{-1}(m_q) - r ],  \hspace{4mm} r = \frac{1}{2m_0}
\EAN
where   
\BAN
D(m_q) = m_q + (m_0 - m_q/2)[ 1 + \gamma_5 S(H_w) ], 
\EAN
thus the quark propagator can be obtained by solving 
the system $ D(m_q) Y = \Id $ with nested conjugate 
gradient \cite{Neuberger:1998my},
which turns out to be highly efficient (in terms of the precision 
of chirality versus CPU time and memory storage) if the inner 
conjugate gradient loop is iterated with Neuberger's double pass 
algorithm \cite{Neuberger:1998jk}.
For more details of our scheme of computing quark propagators, 
see Ref. \cite{Chiu:2003iw}.

We generate 100 gauge configurations with Wilson gauge action
at $ \beta = 6.1 $ on the $ 20^3 \times 40 $ lattice.
Then we compute two sets of (point-to-point) quark propagators,
for periodic and antiperiodic boundary conditions in the time
direction respectively. Here the boundary condition
in any spatial direction is always periodic.
Now we use the averaged quark propagator to compute the
time correlation function for any hadronic observable such that
the effects due to finite $ T $ can be largely
reduced \cite{Chiu:2005is}.

Fixing $ m_0 = 1.3 $, we project out 16 low-lying eigenmodes of 
$ |H_w| $ and perform the nested conjugate gradient in the complement
of the vector space spanned by these eigenmodes. For  
$ N_s = 128 $, 
the weights $ \{ \omega_s \} $ are fixed with $ \lambda_{min} = 0.18 $ 
and $ \lambda_{max} = 6.3 $, 
where $ \lambda_{min} \le \lambda(|H_w|) \le \lambda_{max} $
for all gauge configurations.    

For each configuration, (point-to-point) quark propagators are computed
for 30 bare quark masses in the range $ 0.03 \le m_q a \le 0.8 $, 
with stopping criteria $ 10^{-11} $ and $ 2 \times 10^{-12} $
for the outer and inner conjugate gradient loops respectively.
Then the chiral symmetry breaking due to finite $ N_s (=128) $ is 
less than $ 10^{-14} $,
\BAN
\sigma = \left| \frac{Y^{\dagger} S^2 Y}{Y^{\dagger} Y} - 1 \right|
< 10^{-14},
\EAN
for every iteration of the nested conjugate gradient, and  
the norm of the residual vector for each column of the 
quark propagator is less than $ 2 \times 10^{-11} $  
\BAN
|| (D_c + m_q ) Y - \Id || < 2 \times 10^{-11}  
\EAN

\section{Determination of $ a^{-1} $ and $ m_s $}

After the quark propagators have been computed, we first 
measure the pion propagator and its time correlation function, and 
extract the pion mass ($ m_\pi a $) and the pion decay constant  
($ f_\pi a $). With the experimental input $ f_\pi = 132 $ MeV,
we determine $ a^{-1} = 2.237(76) $ GeV.

The bare mass of strange quark is determined by extracting the
mass of vector meson from the time correlation function
\BAN
C_V (t) = \frac{1}{3} \sum_{\mu=1}^3 \sum_{\vec{x}}
\tr\{ \gamma_\mu (D_c + m_q)^{-1}_{x,0} \gamma_\mu
     (D_c + m_q)^{-1}_{0,x} \}
\EAN
At $ m_q a = 0.08 $, $ M_V a = 0.4601(44) $,
which gives $ M_V = 1029(10) $ MeV, in good agreement with
the mass of $ \phi(1020) $. Thus we take the strange quark
bare mass to be $ m_s a = 0.08 $.
Then we have 10 quark masses smaller than $ m_s $, i.e.,
$ m_u a = 0.03, 0.035, 0.04, 0.045, 0.05, 0.055, 0.06, 0.065, 0.07, 0.075 $.
In this paper, we work in the isospin limit $ m_u = m_d $.

\begin{figure}[htb]
\begin{center}
\begin{tabular}{@{}cc@{}}
\includegraphics*[height=9cm,width=7cm]{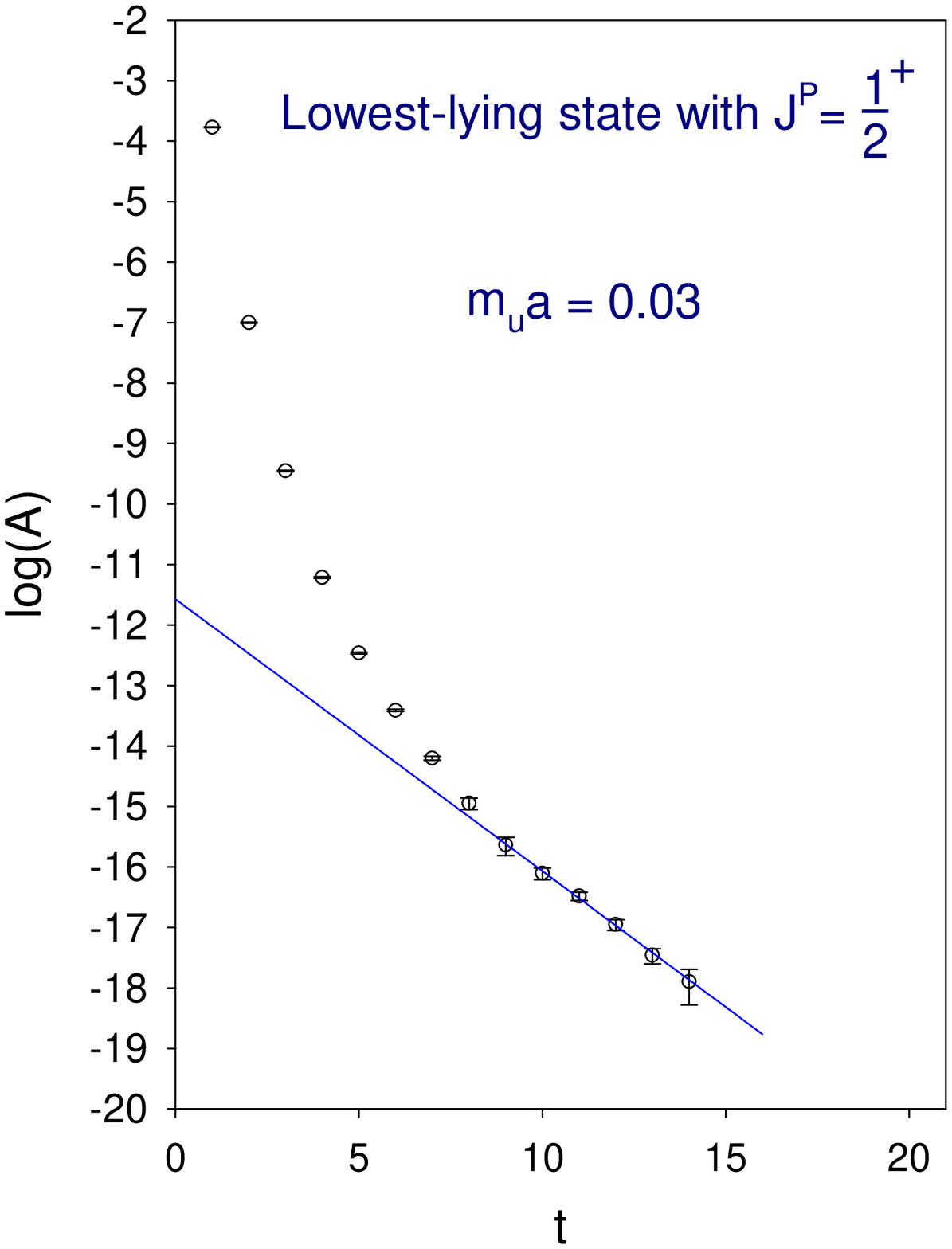}
&
\includegraphics*[height=9cm,width=7cm]{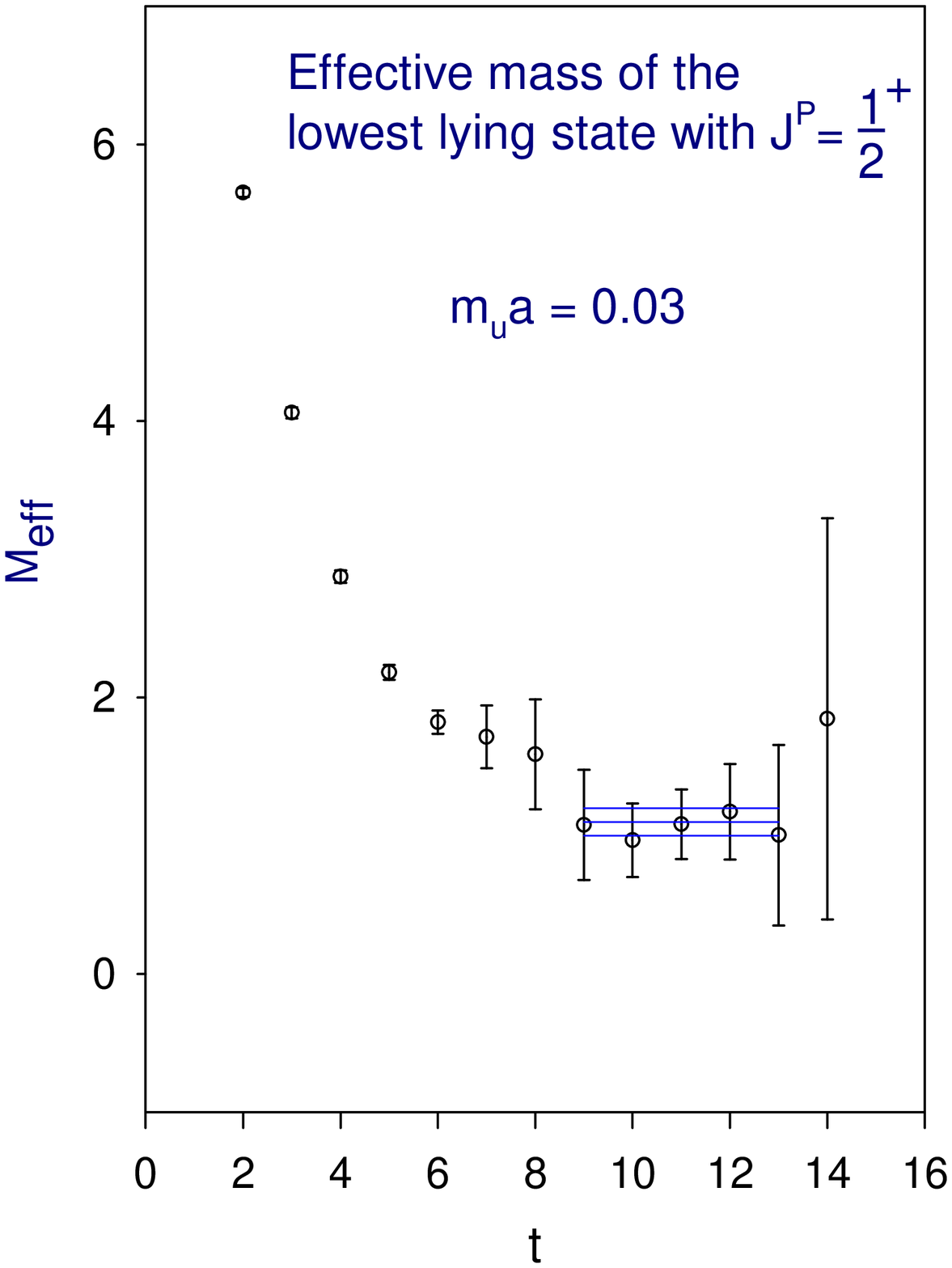}
\\ (a) & (b)
\end{tabular}
\caption{(color online) 
(a) The eigenvalue $ A^+(t) $ of the lowest positive parity state,
for $ m_u a = 0.03 $.
The solid line is the single exponential fit for $ 9 \le t \le 13 $.
(b) The effective mass $ M_{eff}(t) = \ln [A(t)/A(t+1)]  $
of $ A^+(t) $ in Fig.\ \ref{fig:APe3m03}a.
}
\label{fig:APe3m03}
\end{center}
\end{figure}

\begin{figure}[htb]
\begin{center}
\begin{tabular}{@{}cc@{}}
\includegraphics*[height=9cm,width=7cm]{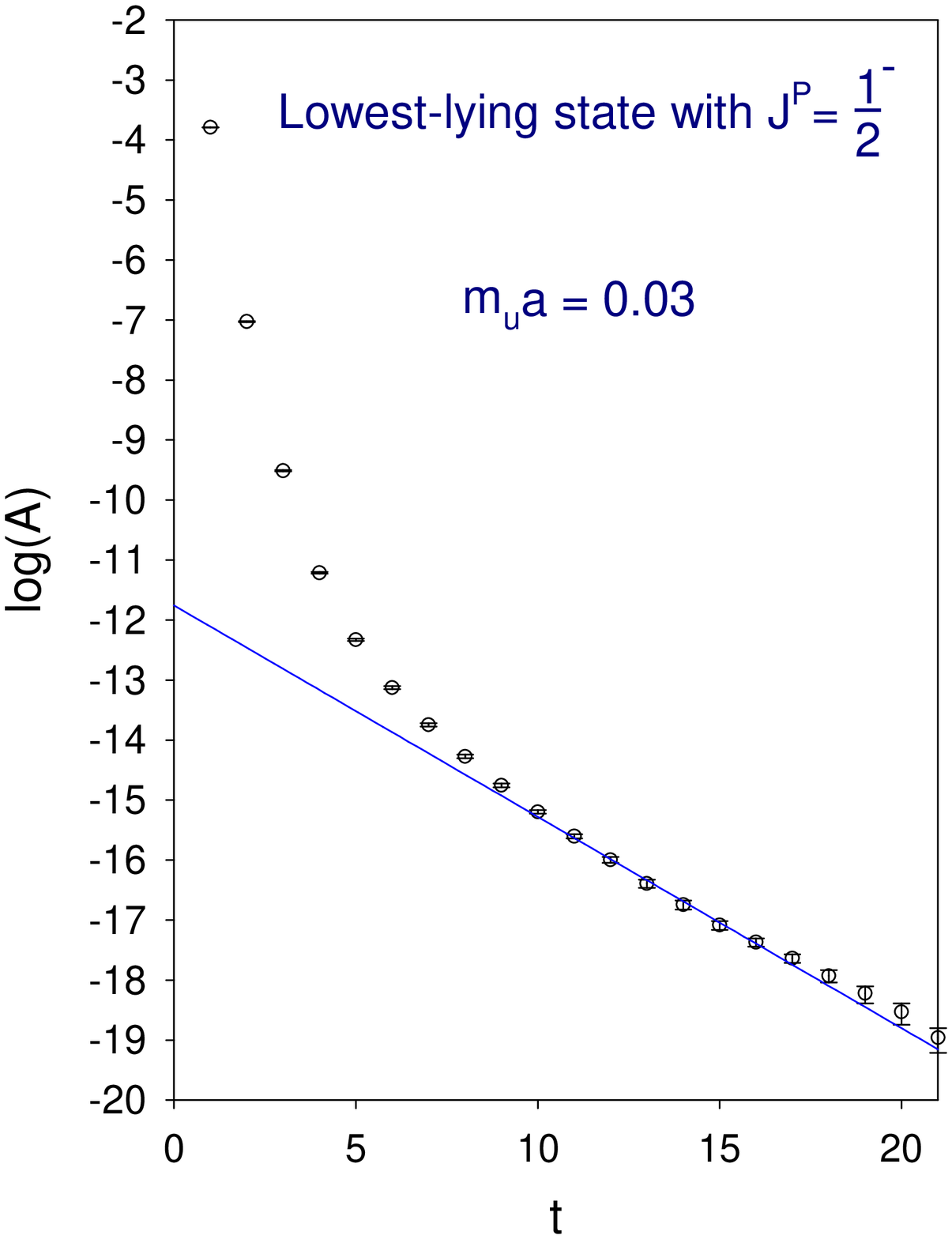}
&
\includegraphics*[height=9cm,width=7cm]{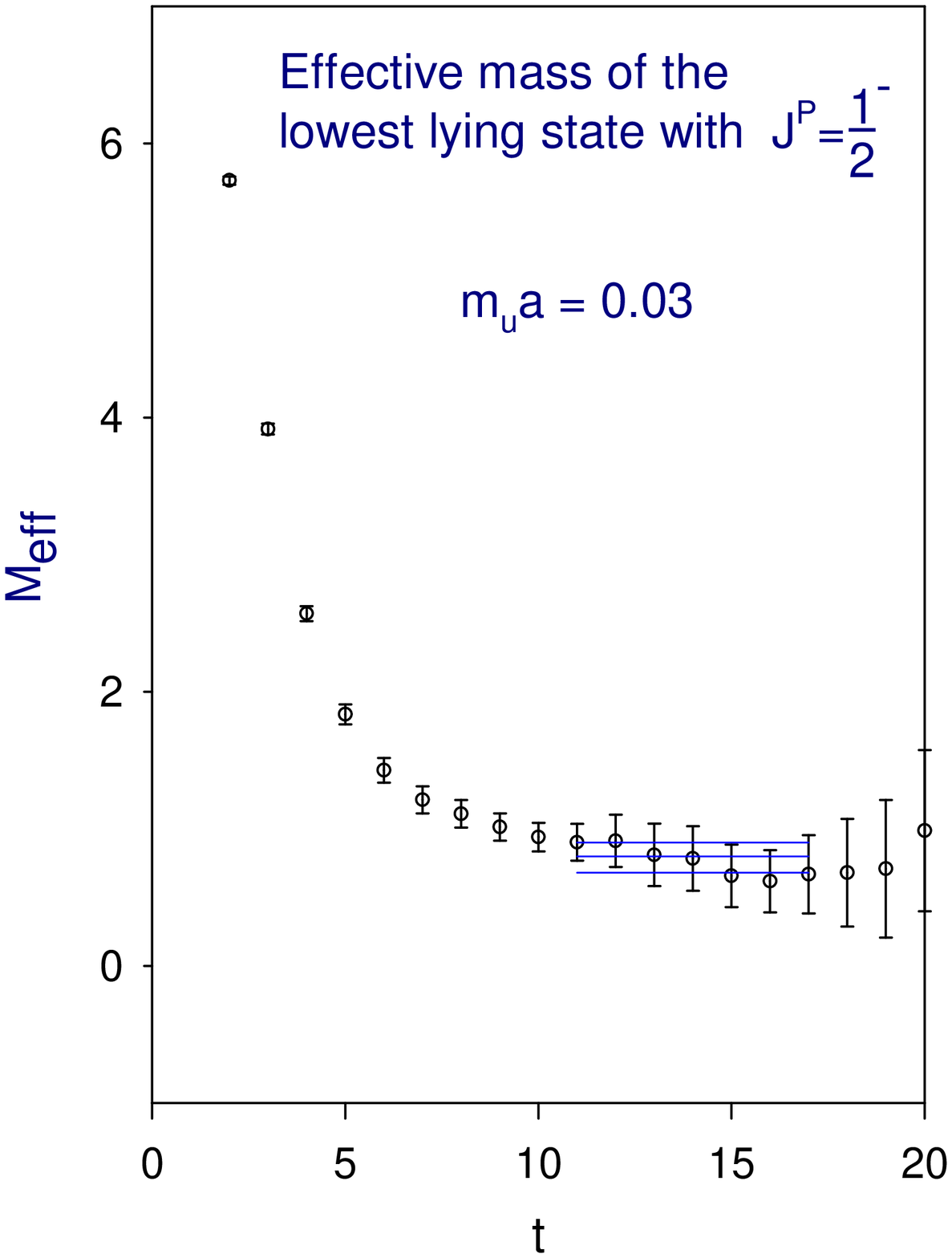}
\\ (a) & (b)
\end{tabular}
\caption{(color online) 
(a) The eigenvalue $ A^-(t) $ of the lowest
negative parity state, for $ m_u a = 0.03 $.
The solid line is the single exponential fit for $ 11 \le t \le 17 $.
(b) The effective mass of $ A^{-}(t) $ in Fig.\ \ref{fig:AMe1m03}a.}
\label{fig:AMe1m03}
\end{center}
\end{figure}

\begin{figure}[htb]
\begin{center}
\begin{tabular}{@{}cc@{}}
\includegraphics*[height=9cm,width=7cm]{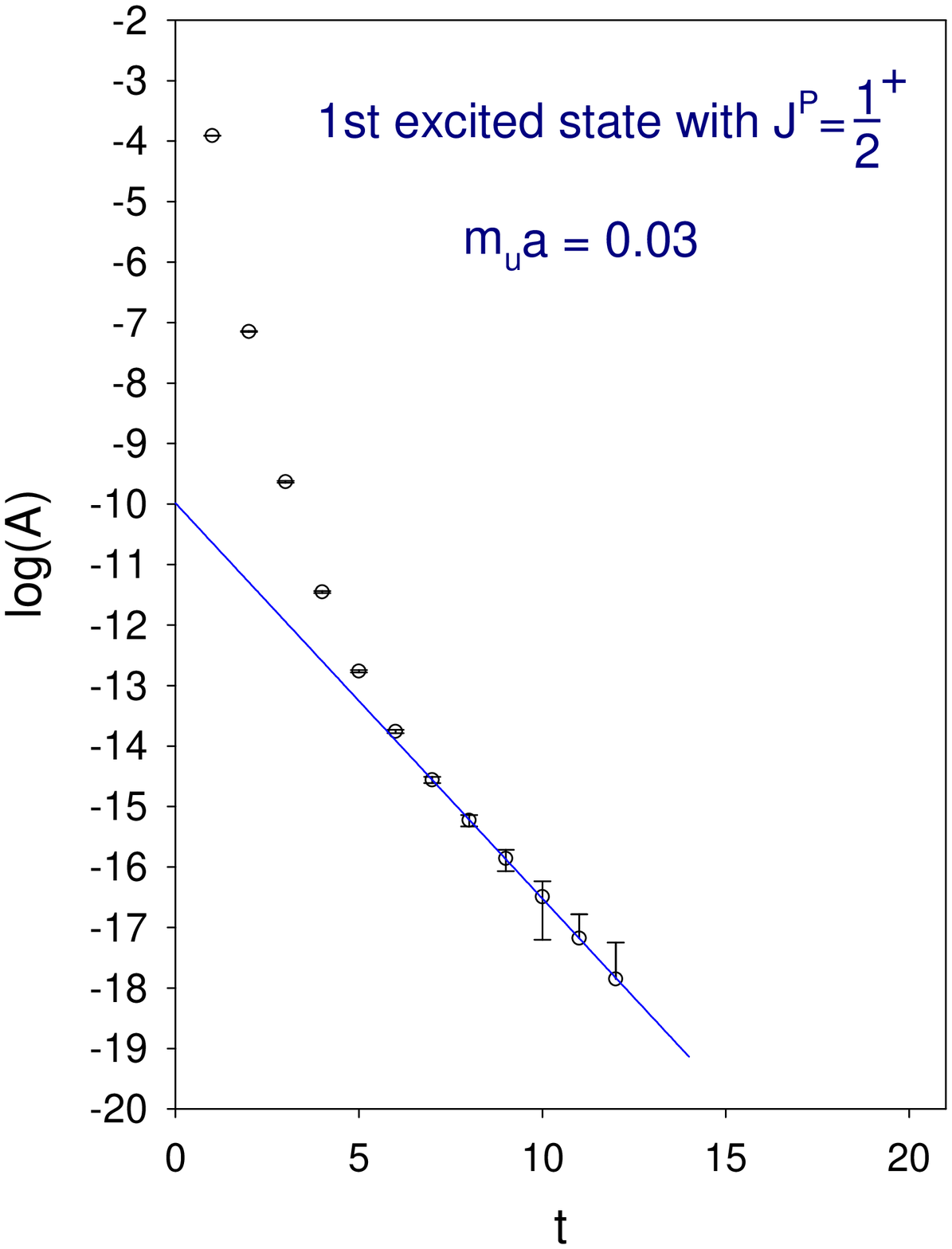}
&
\includegraphics*[height=9cm,width=7cm]{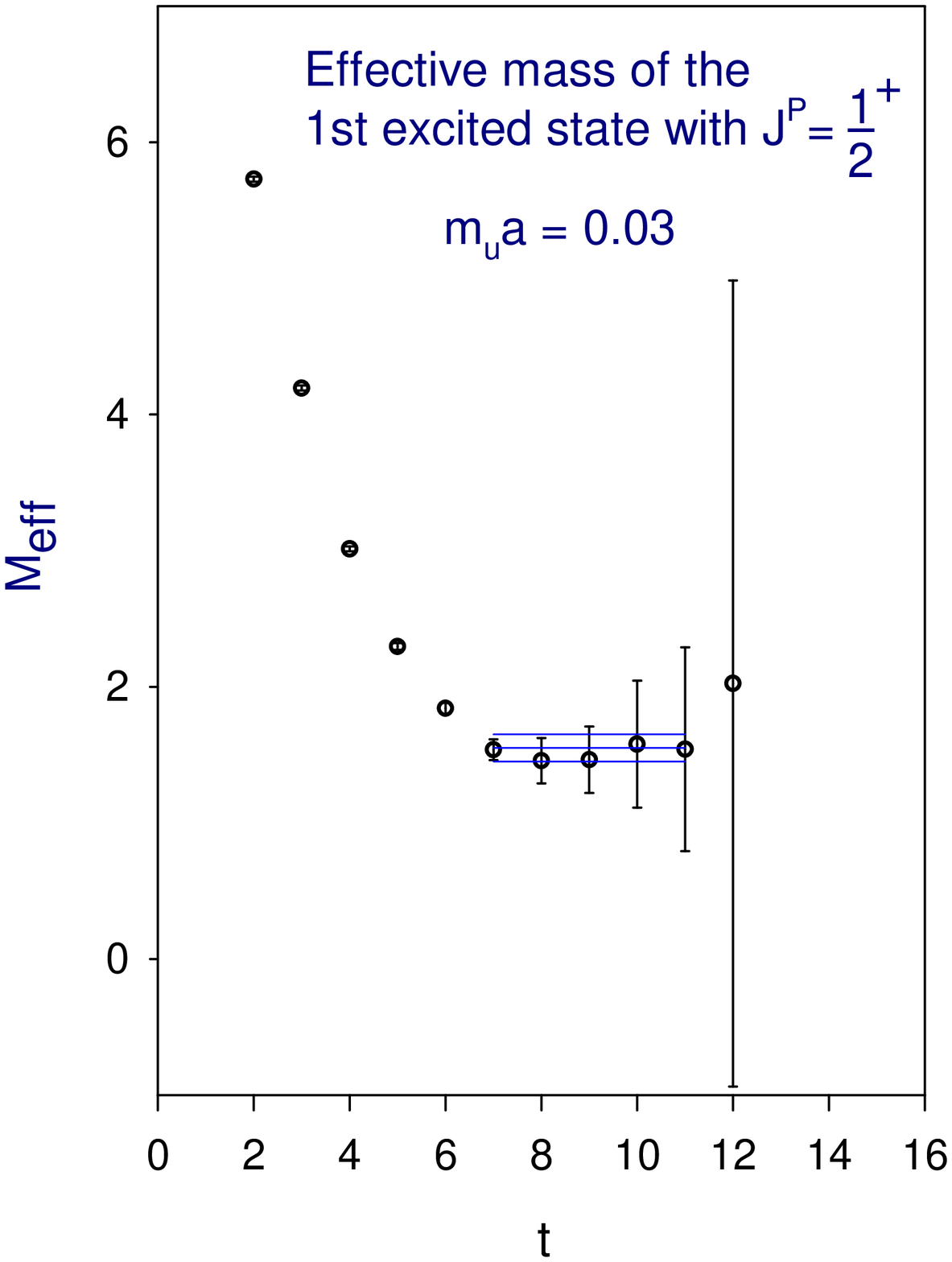}
\\ (a) & (b)
\end{tabular}
\caption{(color online) 
(a) The eigenvalue $ A^+(t) $ of the first excited state
with $ J^P = 1/2^+ $, for $ m_u a = 0.03 $.
The solid line is the single exponential fit for $ 7 \le t \le 11 $.
(b) The effective mass of $ A^{+}(t) $ in Fig.\ \ref{fig:APe2m03}a.}
\label{fig:APe2m03}
\end{center}
\end{figure}

\begin{figure}[htb]
\begin{center}
\begin{tabular}{@{}cc@{}}
\includegraphics*[height=9cm,width=7cm]{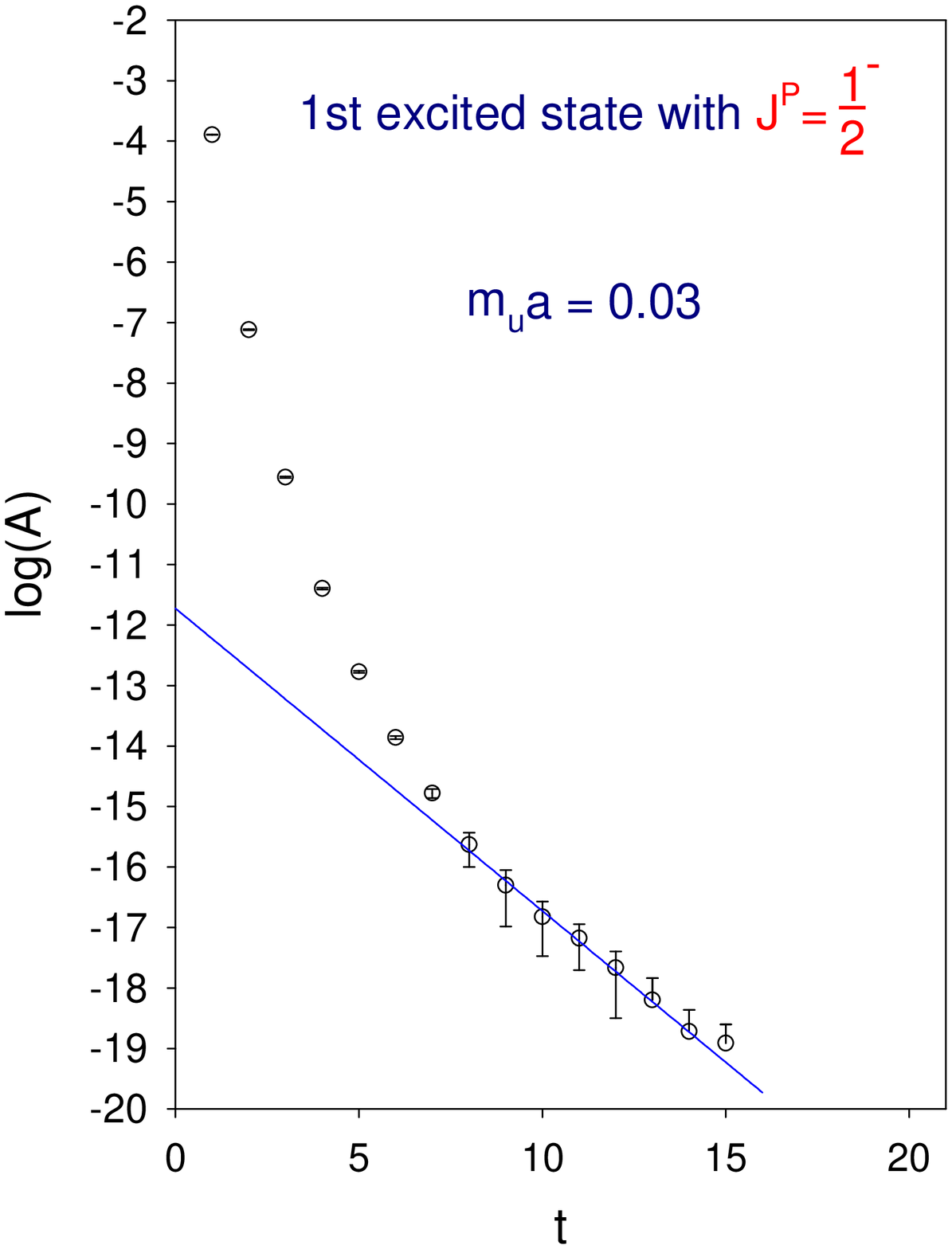}
&
\includegraphics*[height=9cm,width=7cm]{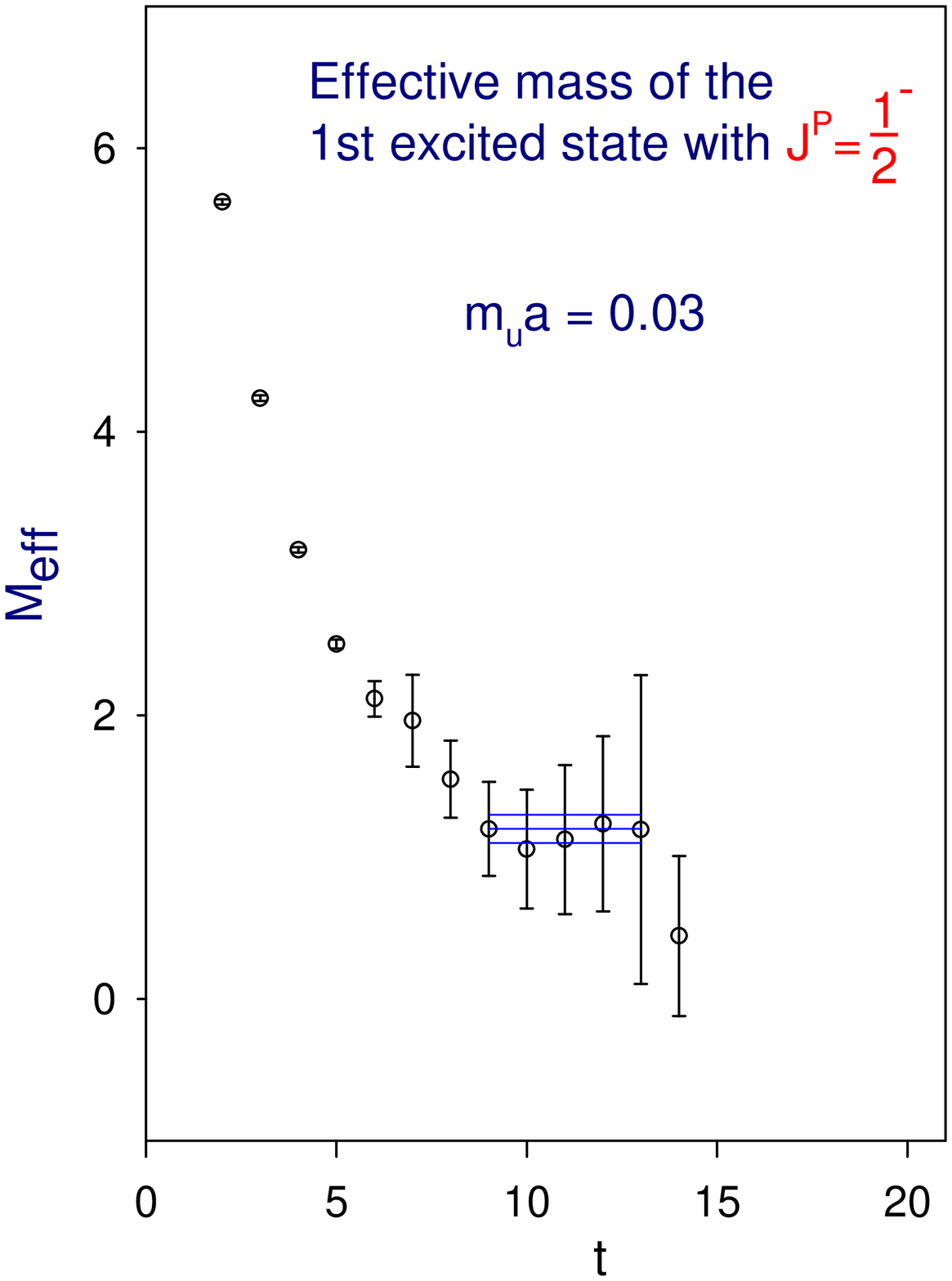}
\\ (a) & (b)
\end{tabular}
\caption{(color online) 
(a) The eigenvalue $ A^-(t) $ of the first excited state
with $ J^P = 1/2^- $, for $ m_u a = 0.03 $.
The solid line is the single exponential fit for $ 9 \le t \le 13 $.
Note that the error becomes very large for $ t > 13 $.
(b) The effective mass of $ A^{-}(t) $ in Fig.\ \ref{fig:AMe2m03}a.}
\label{fig:AMe2m03} \end{center} 
\end{figure}

\section{The $ 3 \times 3 $ correlation matrix for $ \Theta(udud \bar s) $}

Next we compute the propagators
$ \langle (O_i)_{x\alpha} (\bar O_j)_{y\delta} \rangle $
with fixed $ y = (\vec{0},0) $,
and their time correlation functions $ C_{ij}^{\pm}(t) $ with
$ \pm $ parity
\BAN
C_{ij}^{\pm}(t) = \left< \sum_{\vec{x}} \tr \left[ \frac{1\pm\gamma_4}{2}
\langle O_i(\vec{x},t) {\bar O}_j(\vec{0},0) \rangle_f \right] \right>_U
\EAN
where the trace sums over the Dirac space,
and the subscripts $ f $ and $ U $
denote fermionic average and gauge field ensemble average respectively.
Then the $ 3 \times 3 $ correlation matrix
$ C^{\pm}(t) = \{ C_{ij}^{\pm}(t) \} $ can be constructed.
Now, with the variational parameter $ t_0 $, we diagonalize the normalized
correlation matrix $ C^{\pm}(t_0)^{-1/2} C^{\pm}(t) C^{\pm}(t_0)^{-1/2} $
and obtain eigenvalues $ \{ A_i^{\pm}(t) \} $,
and from which to extract the masses $ \{ m_i^{\pm} \} $ of
the lowest-lying and two excited states for $ \pm $ parity respectively.
In general, by varying $ t_0 $, one could minimize the errors 
of the masses extracted from the eigenvalues, as well as to disentangle 
(optimally) the lowest-lying states from the excited ones, 
as shown in Refs. \cite{Michael:1985ne,Luscher:1990ck}.
However, in this case, the relevant quantities (e.g., the effective masses)
extracted from unnormalized correlation matrix seem to be as good as
those of the normalized ones. Thus we restrict to unnormalized
$ C^{\pm}(t) $ in the following. Then the mass $ m_i^{\pm} $ can be
extracted by single exponential fit to $ A_i^{\pm}(t) $,
for the range of $ t $ in which the effective mass
$ M_{eff}(t) = \ln [A(t)/A(t+1)] $ attains a plateau.

In Figs. \ref{fig:APe3m03} - \ref{fig:AMe2m03}, the eigenvalues
$ A^\pm (t) $ corresponding to the lowest-lying and first excited
states with $ J^P = 1/2^\pm $ are plotted versus the time slices, 
for $ m_u a = 0.03 $ (the smallest quark mass in this study), 
together with their effective mass plots. 
Here we have suppressed any data point which has error
(jackknife with single elimination) larger than
its mean value. In each case, the mass $ m^{\pm} $ can be
extracted by single exponential fit to $ A^{\pm}(t) $,
for the range of $ t $ in which the effective mass
$ M_{eff}(t) = \ln [A(t)/A(t+1)] $ attains a plateau. 
The results are (in units of $ a^{-1} $):
\BAN
m(1/2^+)_{\mbox{lowest-lying}}\EQ 1.034(80) \\
m(1/2^+)_{\mbox{1st excited}} \EQ 1.505(137) \\
m(1/2^-)_{\mbox{lowest-lying}}\EQ 0.8045(23) \\
m(1/2^-)_{\mbox{1st excited}} \EQ 1.190(388) 
\EAN
where all fits have confidence level greater than 0.6 and 
$ \chi^2/d.o.f. < 1 $.  
Obviously, the lowest-lying and the first excited states are
disentangled in both parity channels.

\begin{figure}[htb]
\begin{center}
\includegraphics*[height=12cm,width=10cm]{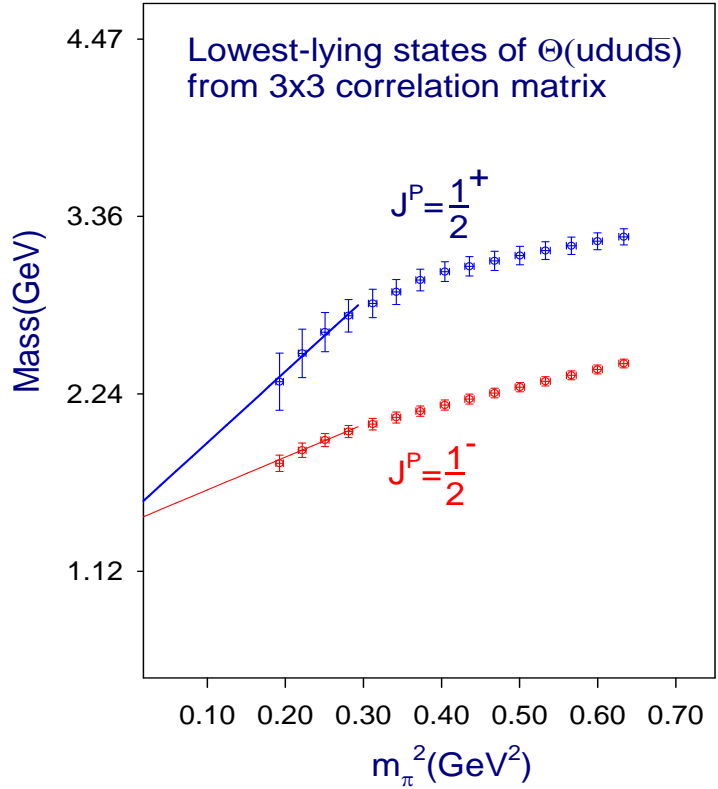}
\caption{(color online)
The masses of the lowest lying states of $ \Theta(udud \bar s) $, 
extracted from the eigenvalues of the $ 3 \times 3 $ 
correlation matrix of $ O_1 $, $ O_2 $ and $ O_3 $.
The solid lines are chiral extrapolation (linear in $ m_\pi^2 $)
using the smallest four masses.}
\label{fig:mPe3Me1}
\end{center}
\end{figure}

In Fig.\ \ref{fig:mPe3Me1}, the masses of the lowest-lying states 
with $ J^P=1/2^\pm $ are plotted versus $ m_\pi^2 $. 
An interesting feature emerges in the positive parity channel. 
Its mass starts to fall more rapidly around the regime 
$ m_u a \simeq 0.045$ (i.e., $ m_u \simeq 0.56 m_s $), 
signaling the onset of certain attractive interactions which lower the 
energy of this pentaquark state. Even taking into account of   
the error bar, the signal is unambiguous, 
as shown in Fig. \ref{fig:APe3m03} 
for $ m_u a = 0.03 $ (the smallest quark mass in this study).   
We suspect that this is the manifestation of diquark
correlations when $ m_{u,d} $ becomes sufficiently small. 
To check, we compare it with 
the positive parity states extracted from the time-correlation 
functions of $ O_1 $, $ O_2 $, and $ O_3 $ respectively 
(see Figs. \ref{fig:mPMo1}-\ref{fig:mPMo3} in the Appendix). Then
we see that similar phenomenon also 
happens in the positive parity channel of $ O_3 $ (Fig. \ref{fig:mPMo3}), 
but {\it not} in $ O_1 $ (Fig. \ref{fig:mPMo1}) 
or $ O_2 $ (Fig. \ref{fig:mPMo2}).
Since $ O_3 $ is the diquark-diquark-antiquark operator, 
it is consistent with our interpretation that  
diquark correlations emerge when $ m_u a \le 0.045 \simeq 0.56 m_s $. 
On the other hand, one may wonder whether it could be an artifact 
due to low statistics. Our argument is that 
if it were due to low statistics, it must also appear 
in the positive parity channel of $ O_1 $ and $ O_2 $.
However, it is not the case. So we rule out the possibility that 
this rapid decrease of the mass of the positive parity state 
in the regime $ m_u a \le 0.045 \simeq 0.56 m_s $ is due to low statistics.  
In other words, it suggests that $ O_3 $ has the largest overlap with 
the pentaquark state, and the diquark correlations may play an important 
role in forming $ \Theta^+ $.

The next question is how to perform the chiral extrapolation 
to the physical limit where $ m_\pi = 135 $ MeV
($ m_{u,d} \simeq m_s/25 $). 
From the viewpoint of chiral perturbation theory, one
should use the set of data points with the smallest $ m_u $ ($ m_\pi^2$). 
Moreover, since our $ m_\pi^{min} $ is about 440 MeV, 
which may not be sufficently small to capture the chiral log behavior 
in chiral perturbation theroy, 
thus we only use the lowest order terms (i.e., linear in $ m_\pi^2 $) 
for chiral extrapolation.   
Observing the onset of diquark correlations around $ m_u a \simeq 0.45 $,  
we naturally pick the smallest four masses (i.e., with
$ m_u a = 0.03, 0.035, 0.04, 0.045 \simeq  0.56 m_s $)
for chiral extrapolation linear in $ m_\pi^2 $. 
At $ m_\pi = 135 $ MeV, we obtain the masses of the lowest-lying states: 
$ m(1/2^-) = 1424(57) $ MeV, and $ m(1/2^+) = 1562(121) $ MeV,
which agree with the masses of
$ m_K + m_N \simeq 1430 $ MeV, and $ \Theta(1540) $ respectively.
For the positive parity state, we have also performed a 
fully correlated fit with the smallest four masses,  
employing the procedure adopted in Ref. \cite{Iwasaki:1995cm}.
Our result is 1554(150) MeV with $ \chi^2_{full}/d.o.f. = 1.17(24) $, 
in agreement with the result of uncorrelated fit. 
By varying the fitting range of $ t $, and the number of mass points, 
we estimate the systematic error to be 180 MeV.

\begin{figure}[htb]
\begin{center}
\includegraphics*[height=12cm,width=10cm]{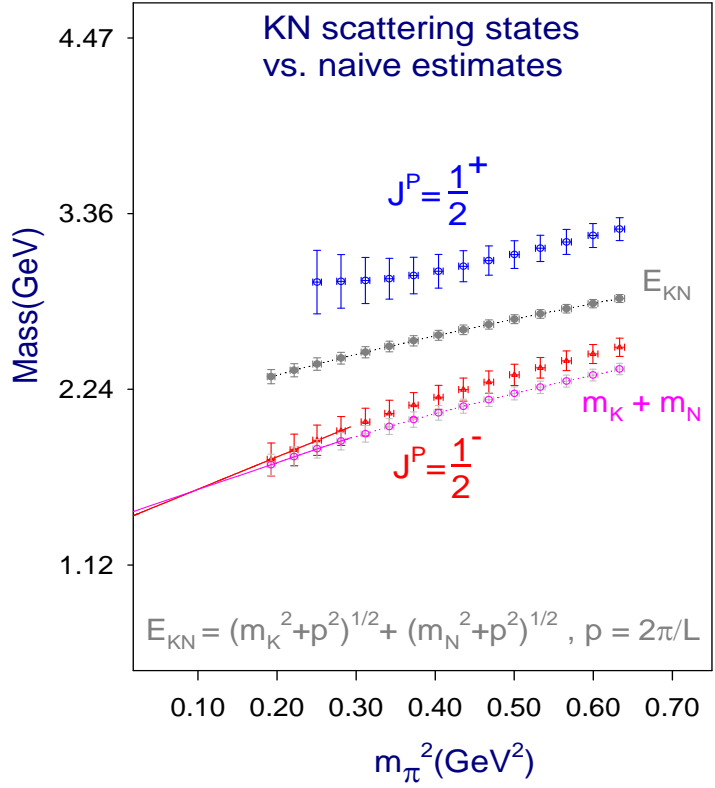}
\caption{(color online)
The masses of the lowest-lying $ KN $ scattering states extracted
from $ C_{KN}(t) $ (\ref{eq:CKN}), versus the naive estimates
(with dotted lines).
The solid lines (for $ J^P = 1/2^- $) are chiral extrapolations
using the smallest four masses.}
\label{fig:mKKNN}
\end{center}
\end{figure}

\begin{figure}[htb]
\begin{center}
\includegraphics*[height=12cm,width=10cm]{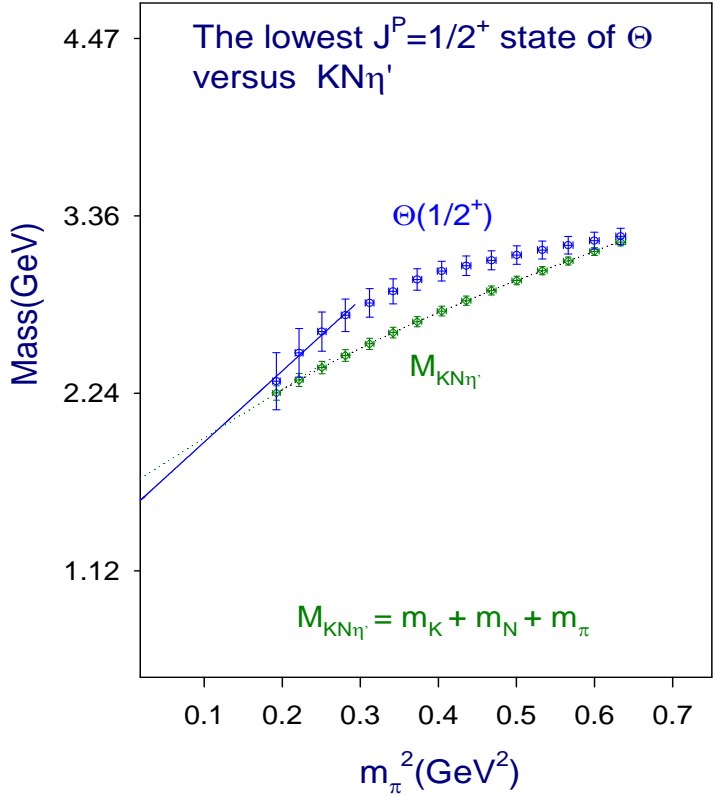}
\caption{(color online)
The mass of the lowest-lying $ J^P = 1/2^+ $ state extracted
from the $ 3 \times 3 $ correlation matrix, versus the s-wave 
of $ KN\eta' $ ghost state (with dotted lines).}
\label{fig:mPe3KNpi}
\end{center}
\end{figure}

\begin{figure}[htb]
\begin{center}
\includegraphics*[height=12cm,width=10cm]{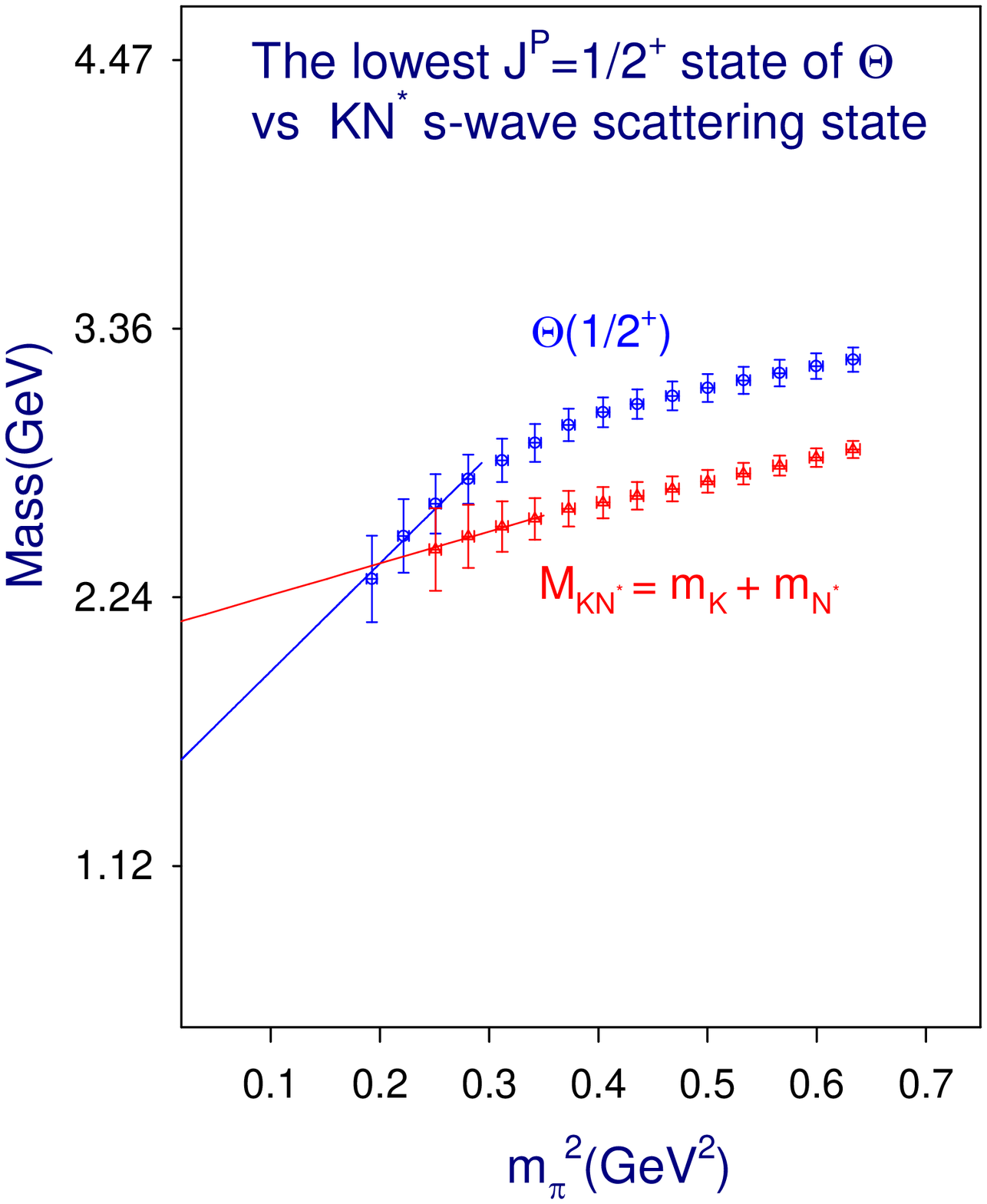}
\caption{(color online)
The mass of the lowest-lying $ J^P = 1/2^+ $ state extracted
from the $ 3 \times 3 $ correlation matrix, versus the  
$ KN^*$ s-wave scattering state with mass $ m_K + m_{N^*} $.}
\label{fig:mPe3_KNS}
\end{center}
\end{figure}

\section{Distinguishing the KN scattering states}

Now the question is whether the lowest-lying $ J^P = 1/2^\pm $ 
states extracted from the $ 3 \times 3 $ correlation matrix
are scattering states or resonances. 
In order to obtain the mass spectrum of $ KN $ 
scattering states (which excludes any pentaquark states), 
we consider the time-correlation function 
of $ KN $ operator without any exchange of quarks between 
$ K $ and $ N $ in its propagator, i.e., the interaction 
between $ K $ and $ N $ is only through the exchange of gluons. 
Explicitly, 
\bea
\label{eq:CKN}
C_{KN}^{\pm}(t) = \left< \sum_{\vec{x}} \tr \left[
                         \frac{1 \mp \gamma_4}{2} 
                  \langle N(\vec{x},t) \bar N(\vec{0},0) \rangle_f
        \langle K(\vec{x},t) \bar K(\vec{0},0) \rangle_f \right] \right>_U
\eea
where $ N=[{\bf u}^T C \gamma_5 {\bf d}] {\bf d} $, and 
$ K=\bar{\bf s}\gamma_5{\bf u} $. Note that the parity projector has 
taken into account of the intrinsic parity of kaon.  

The masses of lowest-lying $ KN $ scattering states are plotted in 
Fig. \ref{fig:mKKNN}, versus the naive estimates. Here the naive
estimates are $ m_K + m_N $ for the s-wave, and 
$ \sqrt{m_K^2 + (2\pi/L)^2} + \sqrt{m_N^2 + (2\pi/L)^2} $ 
for the p-wave, where $ L $ is the lattice size in spatial directions,
and $ m_K $ and $ m_N $ are masses extracted from kaon and nucleon 
time-correlation functions respectively.

For the $ J^P = 1/2^- $ state, 
using the smallest four masses for chiral extrapolation  
to $ m_\pi = 135 $ MeV, we obtain $ m_{KN}(1/2^-) = 1433(72) $ MeV, 
in agreement with $ m_K + m_N \simeq 1430 $ MeV. 
Further, its mass spectrum is almost identical 
to that of the lowest $ J^P = 1/2^- $ state of $ \Theta $ 
in Fig. \ref{fig:mPe3Me1}, for the entire range of $ m_u $. 
Thus we identify the lowest $ J^P = 1/2^- $ state of $ \Theta(udud\bar s) $ 
with the KN s-wave scattering state.

On the other hand, for the $ J^P = 1/2^+ $ state in Fig. \ref{fig:mKKNN},
its mass is higher than the naive estimate 
$ \sqrt{m_K^2 + (2\pi/L)^2} + \sqrt{m_N^2 + (2\pi/L)^2} $. 
This suggests that the KN p-wave (in the quenched approximation)
in a finite torus is more complicated than just two free particles
with momenta $ \vec{p}_K = -\vec{p}_N = 2\pi\hat{e_i}/L $, i.e., 
their interaction through gluon exchanges cannot be ignored. 
Further, the mass of KN p-wave scattering state in Fig. \ref{fig:mKKNN}
is always higher than the mass of the lowest-lying $ J^P = 1/2^+ $ state 
in Fig. \ref{fig:mPe3Me1}. In particular, for $ m_u < m_s $, 
the former is significantly higher than the latter. 
This seems to suggest that the lowest $ J^P = 1/2^+ $ state of 
$ \Theta(udud \bar s) $ is {\it different} from the KN p-wave 
scattering state. 

However, there are other two hadron scattering states which also 
have quantum numbers $ J^P = 1/2^+ $ and $ S=+1 $, namely, the s-wave 
scattering state of $ KN\eta'$ (where $ \eta' $ is an artifact due to 
quenched approximation), and the s-wave scattering state of $ K N^* $ 
(where $ N^* $ is the negative parity state of nucleon). 
In the following, we check whether the lowest-lying $ J^P = 1/2^+ $ 
state extracted from the $ 3 \times 3 $ correlation matrix could 
possibily be any one of these two hadron scattering states.

The s-wave of $ KN\eta'$ has $ J^P = 1/2^+ $, and its mass 
is estimated to be $ m_K + m_N + m_\pi $ \cite{Mathur:2004jr}.
In Fig. \ref{fig:mPe3KNpi}, it is clear that the lowest-lying 
$ J^P = 1/2^+ $ state of $ \Theta $ is {\it different} 
from the s-wave of $ KN\eta' $ ghost state. Otherwise,  
they should almost identical for the entire range of $ m_u $, 
as in the case of $ KN $ s-wave and the lowest-lying state 
with $ J^P = 1/2^- $ in Fig. \ref{fig:mKKNN}. 
So we exclude the possibility that the lowest-lying $ J^P=1/2^+ $ state 
extracted from the $ 3 \times 3 $ correlation matrix is due to 
quenched artifacts. Note that Mathur et. al. \cite{Mathur:2004jr}
has claimed that the quenched artifact $ KN \eta' $ 
(with negative spectral weight) would be seen in the positive parity 
channel if $ m_\pi < 300 $ MeV, for the  
$ O_1 $ ($ K \otimes \gamma_5 N $) operator.  
If this claim holds for any cases 
(gauge actions, lattice fermions, operators, etc),
then we should not see $ KN \eta' $, since our smallest
pion mass is larger than 400 MeV.

Next, we turn to the s-wave of $ KN^*$, which
also has $ J^P = 1/2^+ $, and its mass is estimated to be
$ m_K + m_{N^*} $. In Fig. \ref{fig:mPe3_KNS}, it is clear that
the mass of $ K N^* $ s-wave is {\it different} from
the mass of the lowest $ J^P = 1/2^+ $ state
extracted from the $ 3 \times 3 $ correlation matrix,  
for the entire range of $ m_u $ (except at the crossover 
$ m_u a \simeq 0.03 $). In the physical pion limit, 
the mass of the lowest $ J^P = 1/2^+ $ state is $ 1562(121) $ MeV, 
while $ m_K + m_{N^*} $ is $ 2137(142) $ MeV.  
This rules out the possibility that the lowest-lying $ J^P=1/2^+ $ 
state extracted from the $ 3 \times 3 $ correlation matrix is 
the $ KN^* $ s-wave scattering state. 
Note that this disagrees with the claim of   
Ref. \cite{Takahashi:2004sc}, which was based on their result 
obtained with Wilson quarks at unphysical quark masses. 
However, there is no reason to expect that baryon masses computed
with two different lattice (fermion and gauge) actions would agree with
each other, except in the continuum (and infinite volume) limit 
with physical quark masses. Thus, it is {\it not} surprising 
to see that, at unphysically large quark masses, our result of the mass 
of the lowest-lying state with even parity, $ m(1/2^+) $, is higher than 
$ m_K + m_{N^*} $, while the result of Ref. \cite{Takahashi:2004sc} 
seems to suggest that $ m(1/2^+) \simeq m_K + m_{N^*} $.
Now the emerging problem for Wilson fermion is to check 
whether the relationship $ m(1/2^+) \simeq m_K + m_{N^*} $ remains valid 
in the physical pion limit. We suspect that even for Wilson fermion, 
it would also exhibit the inequality $ m(1/2^+) < m_K + m_{N*} $ 
at sufficiently small quark masses, though the location of 
the crossover might be different from that of Fig. \ref{fig:mPe3_KNS}.  
At this point, it is instructive to check the result of 
Mathur et al. \cite{Mathur:2004jr}, which was obtained with the 
Overlap fermion. From Ref. \cite{Mathur:2004jr} and hep-ph/0306199,   
one can extract $ m(1/2^+) $, $ m_K $, and $ m_{N^*} $ respectively,   
then one immediately sees that $ m(1/2^+) $ is larger than $ m_K + m_{N^*} $ 
for $ m_\pi^2 > 0.27 \mbox{ GeV}^2 $, but becomes  
smaller than $ m_K + m_{N^*} $ for $ m_\pi^2 < 0.27 \mbox{ GeV}^2 $,    
and it tends to $ \sim 1650 $ MeV in the physical pion limit.
Obviously, the result of Mathur et al. \cite{Mathur:2004jr} also suggests 
that the lowest state with even parity is different from the $ KN^* $ s-wave, 
in agreement with our conclusion.
Now a (physically irrelevant) question is where the 
$ KN^* $ s-wave scattering state lies, say, at unphysically 
heavy quark masses. 
Our conjecture is that in a finite torus, the $ KN^* $ s-wave 
might turn out to be much heavier than the naive estimate $ m_K + m_{N^*} $  
(similar to the case of $ KN $ p-wave which is much heavier 
than its naive estimate, as shown in Fig. \ref{fig:mKKNN}),  
thus it always lies above the lowest 
$ J^P = 1/2^+ $ state, as one of the excited states in the 
positive parity channel.

Now, after ruling out the possibilities of being 
$ KN $ p-wave, $ KN \eta' $ s-wave, or $ KN^* $ s-wave,
the lowest-lying $ J^P = 1/2^+ $ state extracted from the 
$ 3 \times 3 $ correlation matrix seems to be nothing 
but a resonance. 
If it is identified with $ \Theta^+(1540) $, then it
predicts that the parity of $ \Theta^+(1540) $ is positive.

\section{Discussions and concluding remarks}

Now we return to Table \ref{tab:Theta_LQCD_0205} to discuss 
what causes the different claims in these exploratory 
lattice studies. 

Now, back to Fig. \ref{fig:mPe3Me1}. If we had not measured
any data points with $ m_u < m_s/2 $, then we could not have 
seen the rapid decrease of mass of the positive parity state 
(in the regime $ m_u a \le 0.045 $), of which we interpret 
as the manifestation of diquark correlations at sufficently 
small $ m_{u,d} $. Consequently, chiral extrapolations
using data points with $ m_u > m_s /2 $ must yield much
higher masses, especially for the positive parity state.  
For example, if we use the data points with
$ m_u a = 0.05, 0.055, 0.06, 0.065 < m_s a $, then we obtain
$ m(1/2^-) = 1666(47) $ MeV, and $ m(1/2^+) = 2178(104) $ MeV,
with chiral extrapolation linear in $ m_\pi^2 $.
Further, if we use the data points with
$ m_u a = 0.07, 0.075, 0.08, 0.085 (> m_s a) $, then we obtain
$ m(1/2^-) = 1725(52) $ MeV, and $ m(1/2^+) = 2616(103) $ MeV.
In either one of these two cases, it seemingly rules out the possibility
that the positive parity channel could accommodate any state
with mass $ 1540 $ MeV. However, it is only {an artifact due to
chiral extrapolation with data points too far away from the physical
reality} ($ m_{u,d} \simeq m_s/25 $).
This explains why the masses of the positive parity
state in Refs. \cite{Sasaki:2003gi,Ishii:2004qe}
are so high comparing to our result.
Note that, in Refs. \cite{Sasaki:2003gi, Ishii:2004qe}, 
the total number of data points is four, in which the number of
``physical" data points (i.e. with $ m_u < m_s $) is only one and zero
(see Table \ref{tab:Theta_LQCD_0205}).
Even though they claimed that the positive parity channel could not
accommodate $ \Theta^+(1540) $, it is most likely just an artifact
due to their chiral extrapolations with data points
too far away from the physical reality.

\begin{figure}[htb]
\begin{center}
\includegraphics*[height=12cm,width=10cm]{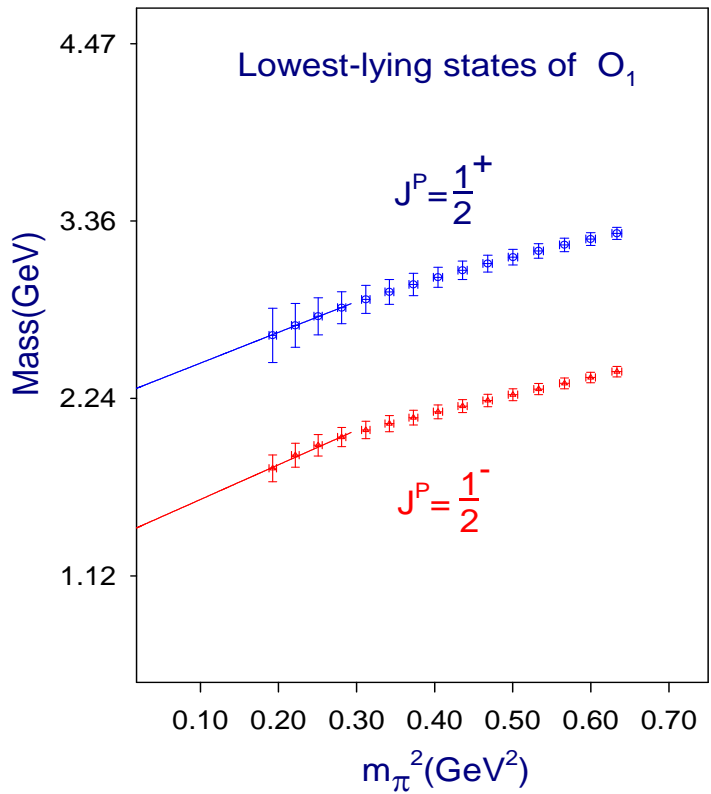}
\caption{(color online)
The masses of the even and odd parity states extracted 
from the time-correlation function of $ O_1 $.
The solid lines are chiral extrapolation (linear in $ m_\pi^2 $)
using the smallest four masses.}
\label{fig:mPMo1}
\end{center}
\end{figure}

\begin{figure}[htb]
\begin{center}
\includegraphics*[height=12cm,width=10cm]{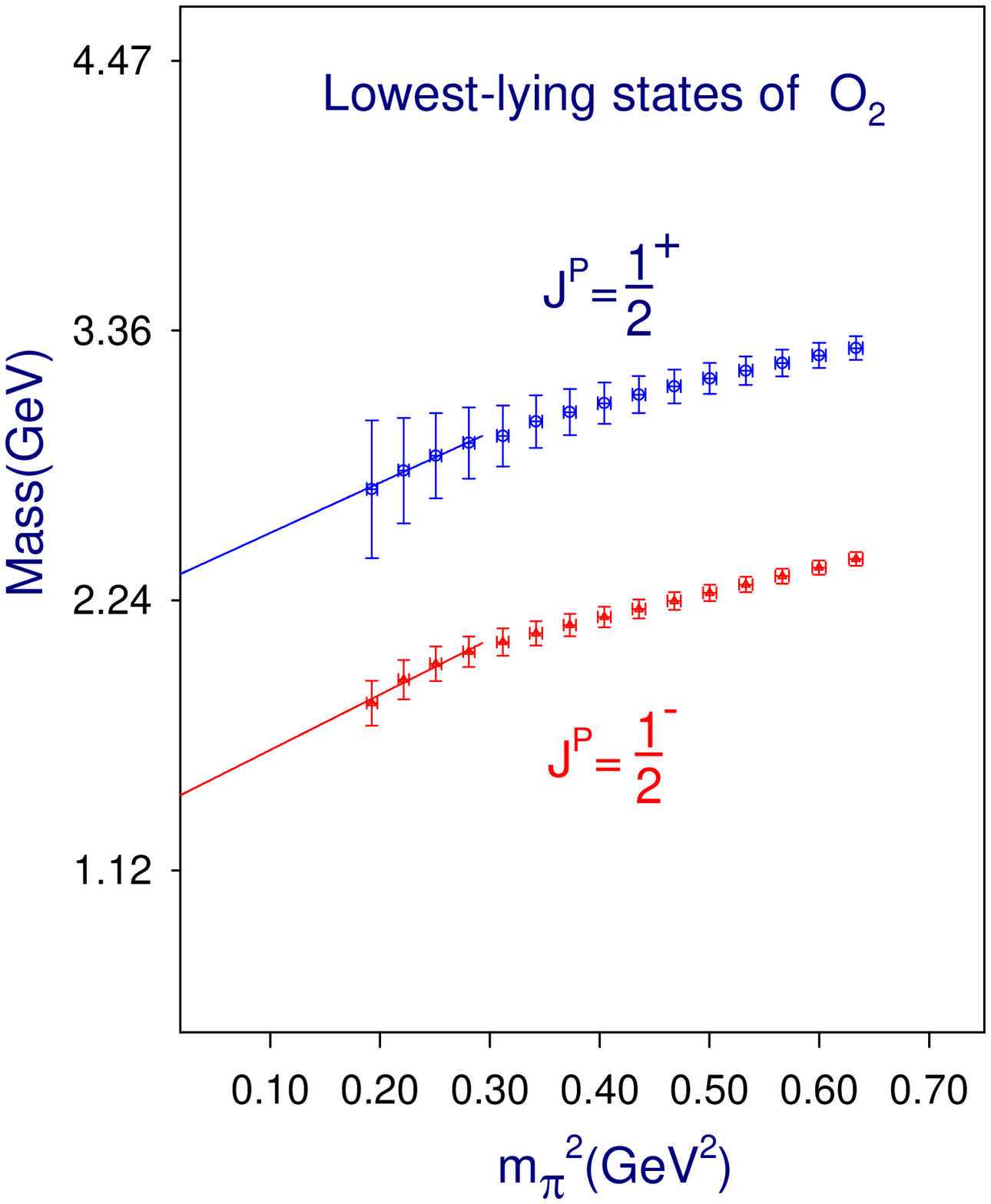}
\caption{(color online)
The masses of the even and odd parity states extracted 
from the time-correlation function of $ O_2 $.
The solid lines are chiral extrapolation (linear in $ m_\pi^2 $)
using the smallest four masses.}
\label{fig:mPMo2}
\end{center}
\end{figure}

Next, we turn to the claims of Refs. \cite{Csikor:2003ng,Mathur:2004jr}.
From our mass spectra of $ O_1 $ (Fig. \ref{fig:mPMo1})
and $ O_2 $ (Fig. \ref{fig:mPMo2}), one can explain why 
Csikor et al. \cite{Csikor:2003ng} (with $ O_1 $ and $ O_2 $)
obtained a rather high mass for the positive parity state,
while Mathur et al. \cite{Mathur:2004jr} (with $ O_1 $)
did not see pentaquark resonance in the positive parity channel.
Namely, the interpolating operators $ O_1 $ and $ O_2 $
have little overalp with the pentaquark state in the positive
parity channel. In fact, our mass spectra of $ O_1 $ and $ O_2 $
are consistent with those of Refs. \cite{Csikor:2003ng,Mathur:2004jr}.

Finally we discuss the results of this paper. 
This is the first lattice QCD study on $ \Theta $ with 
$ 3 \times 3 $ correlation matrix. Presumably, it should 
provide a more reliable answer to the questions of signal/parity   
of $ \Theta^+ $ than other lattice studies with only one operator. 
However, this is a quenched lattice QCD calculation (like other lattice
studies on $ \Theta^+ $ so far), with only one volume, 
and one lattice spacing, thus it is difficult for us to estimate the 
systematic error. 
For the lowest $ J^P = 1/2^- $ state, it is identified with 
the $ KN $ s-wave, by comparing its mass with $ m_N + m_K $.
For the lowest $ J^P = 1/2^+ $ state, by comparing its mass 
(as a function of $ m_\pi^2 $) to those of two-hadron scattering states 
having the same quantum numbers, it seems unlikely to be identified with 
any one of the following two-hadron scattering states: 
$ KN $ p-wave, $ KN \eta' $ s-wave, and $ KN^* $ s-wave.   
However, before it can be confirmed to be a resonance, 
it is necessary to check whether its mass and spectral weight 
are volume independent. 
To this end, we are performing computations on 
the $ 24^3 \times 48 $ lattice with the same lattice spacing 
(i.e., Wilson gauge action at $ \beta = 6.1 $).  
If the lowest-lying state with $ J^P = 1/2^+ $ turns out to be 
a scattering state, then there is no evidence of pentaquark resonance 
in our study. On the other hand, if it turns out to be 
a resonance, then it can be identifed with $ \Theta^+(1540) $ 
since its mass is close to 1540 MeV. 
Nevertheless, one still has to find out whether its decay 
width could be as small as 15 MeV (compatible to that of 
$ \Theta^+ $), which is the most 
challenging problem pertaining to $ \Theta^+ $.

\bigskip

\flushpar
{\bf Appendix}

\bigskip

\noindent

In this appendix, we present our results of the masses of 
the even and odd parity states extracted from the 
time-correlation functions of operators 
$ O_1 $ (\ref{eq:O1}), $ O_2 $ (\ref{eq:O2}),
and $ O_3 $ (\ref{eq:O3}) respectively.    

\begin{figure}[htb]
\begin{center}
\hspace{0.0cm}\includegraphics*[height=12cm,width=10cm]{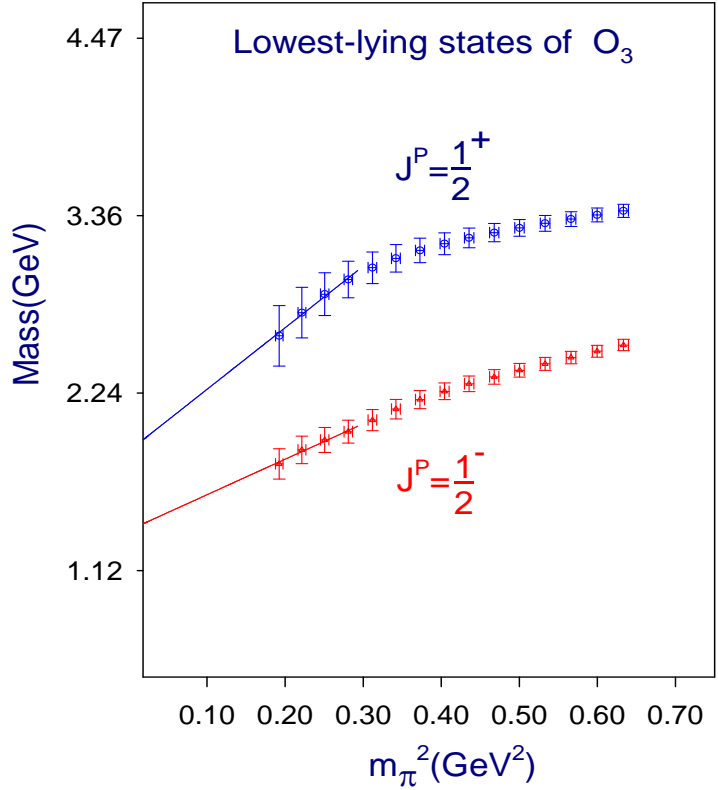}
\caption{(color online)
The masses of $ J^P = 1/2^\pm $ states extracted from the 
time-correlation function of $ O_3 $, 
the diquark-diquark-antiquark operator.
The solid lines are chiral extrapolation (linear in $ m_\pi^2 $)
using the smallest four masses.}
\label{fig:mPMo3}
\end{center}
\end{figure}

In Figs. \ref{fig:mPMo1}-\ref{fig:mPMo3},
the masses of the $ J=1/2^\pm $ states are plotted
versus $ m_\pi^2 $, for $ O_1 $, $ O_2 $, and $ O_3 $ respectively.
Here all mass fits have confidence level greater than 0.6 and 
$ \chi^2/d.o.f. < 1 $.
For $ O_1 $ and $ O_2 $, their masses vary smoothly with respect
to $ m_\pi^2 $, in both parity channels. On the other hand, 
the positive parity state of $ O_3 $ undergoes a 
rapid decrease for $ m_u a \le 0.045 $, similar to the behavior 
of the lowest-lying $ J^P = 1/2^+ $ state extracted from the 
$ 3 \times 3 $ correlation matrix (see Fig. \ref{fig:mPe3Me1}). 
This seems to signal an onset of certain attractive interactions 
when the quark mass $ m_{u,d} $ becomes 
sufficiently small. We conjecture that this is the manifestation 
of diquark correlations when $ m_{u,d} $ approaches the 
physical limit.  
 
Following the same argument in Section 4, we pick the smallest 
four masses (i.e., with $ m_u a = 0.03, 0.035, 0.04, 0.045 $)
for chiral extrapolation (linear in $ m_\pi^2 $). 
At physical pion mass $ m_\pi = 135 $ MeV, we obtain
\BAN
O_1: &&  m(1/2^-) = 1430 (66) \mbox{MeV}, \hspace{4mm} 
         m(1/2^+) = 2301(134) \mbox{MeV}  \\
O_2: &&  m(1/2^-) = 1430 (67) \mbox{MeV}, \hspace{4mm} 
         m(1/2^+) = 2346(156) \mbox{MeV}  \\
O_3: &&  m(1/2^-) = 1446 (71) \mbox{MeV}, \hspace{4mm} 
         m(1/2^+) = 1843(136) \mbox{MeV}
\EAN

Note that for the $ J^P = 1/2^- $ state, all three operators 
give almost the same mass which coincides with that [$1424(57)$ MeV] 
extracted from the $ 3 \times 3 $ correlation matrix of 
$ O_1 $, $ O_2 $, and $ O_3 $. 
However, for the $ J^P = 1/2^+ $ state,        
the operator $ O_3 $ (the diquark-diquark-antiquark operator
motivated by the Jaffe-Wilczek model) gives the lowest mass, 
which is the closest to that [$1562(121)$ MeV] extracted from 
the $ 3 \times 3 $ correlation matrix of $ O_1 $, $ O_2 $, and $ O_3 $. 
This seems to imply that among $ \{ O_i, i=1,2,3 \} $, 
$ O_3 $ has the largest overlap with the pentaquark state, 
and the diquark correlations may play an important 
role in forming $ \Theta^+(1540) $. 

Further, we also observe that the diagonalization of the $ 3 \times 3 $ 
correlation matrix indeed disentangles the contributions of the excited 
states, and gives a smaller mass for the lowest-lying $ J^P = 1/2^+ $ state.

\eject

\flushpar
{\bf Acknowledgement}

\bigskip

\noindent

This work was supported in part by the National Science Council (ROC)
under Grant No. NSC93-2112-M002-016, and 
the National Center for High Performance Computation at Hsinchu, 
and the National Science Foundation (US) under Grant No. PHY99-07949.
This paper was completed while TWC participating 
the Program ``Modern Challenges for Lattice Field Theory" at Kavli 
Institute for Theoretical Physics (KITP), University of California, 
Santa Barbara. TWC would like to thank the Organizers of the Program, 
and KITP for supports and kind hospitality.


\vfill\eject

\end{document}